\documentclass[twocolumn,nofootinbib,showpacs,prd]{revtex4}%
\usepackage{amsmath}
\usepackage{amsfonts}
\usepackage{amssymb}
\usepackage{graphicx}

\def\be{\begin{equation}}
\def\ee{\end{equation}}
\def\bea{\begin{eqnarray}}
\def\eea{\end{eqnarray}}

\begin{document}

\title{Reconstructing the expansion history of the Universe with a one-fluid approach}

\author{Orlando Luongo${}^\dag{}^\ddag{}^\S$ and Hernando Quevedo${}^\dag{}^\S$}\email{orlando.luongo@roma1.infn.it,quevedo@nucleares.unam.mx}
\address{$^\dag$Dipartimento  di Fisica and Icra, Universit\`a di Roma "La Sapienza", Piazzale Aldo Moro 5, I-00185, Roma, Italy;\\
$^\ddag$Dipartimento di Scienze Fisiche, Universit\`a di Napoli "Federico
II", Via Cinthia, I-80126, Napoli, Italy;\\$^\S$Instituto de Ciencias Nucleares, Universidad Nacional Aut\'onoma de M\'exico, AP 70543,
M\'exico DF 04510, Mexico.}

\begin{abstract}
Assuming that the Universe is filled by one single fluid, we present in the context of General Relativity a possible explanation for the acceleration of the Universe. We use ordinary thermodynamics and the fact that small matter perturbations barely propagate in our Universe, to derive a general solution
for a single fluid in which the speed of sound vanishes. We find a model that contains $\Lambda$CDM as a special case, and is compatible with current
observational data.
\end{abstract}


\maketitle

To explain the (positive) acceleration of the Universe
\cite{SNeIa}, it is usually assumed that,
besides a dust-like fluid, the Universe is filled by an
additional exotic fluid that accounts for about 75\% of the energy density in
the Universe. Many models are known in the literature that intend to describe
the nature of this additional component \cite{copeland,au}; the simplest explanation is obtained
by assuming the existence of a cosmological constant which is the basic ingredient
of the well-known $\Lambda$CDM model.
Despite its simplicity, the
 $\Lambda$CDM model  suffers from various theoretical
shortcomings \cite{Lambda}, so it appears inadequate to be
considered as a definitive model.
In this work, we propose a different approach
based on the assumption that the dynamics of the Universe satisfies the laws
of ordinary thermodynamics and on the observational fact that the speed of
sound vanishes for a matter fluid \cite{sound}.

A perfectly homogeneous and isotropic cosmology with zero spatial
curvature\footnote{We limit ourselves to this case to be in accordance with observations.
The generalization to the case of nonzero spatial curvature is straightforward.}
is described by the Friedman-Robertson-Walker (FRW) line element
\be
ds^2=dt^2-a(t)^2(dr^2+r^2\sin^2\theta d\phi^2) \ .
\label{frw}
\ee
Moreover, the gravitational source is assumed to be described by the energy-momentum tensor of
a perfect fluid \cite{other}: $T_{\mu\nu}={\rm diag}(\rho(t), -p(t), - p(t), -p(t))$. One can show that
for a cosmological model satisfying the above symmetry conditions it is possible
to apply the laws of ordinary thermodynamics in a consistent manner \cite{kqs}.
Let us consider a reversible and frictionless universe, satisfying the first law of thermodynamics,
$dQ=dU+p dV$,
and the equation of state (EoS) of an ideal gas
$p=\rho (C_p-C_V)T$,
where the heat capacities are defined as
$C_V=\left(\frac{\partial U}{\partial T}\right)_{V}$ and
$C_p=\left(\frac{\partial h}{\partial T}\right)_{p}$, with
$h=U+pV$ being the enthalpy of the system. Furthermore, we assume that the evolution
of the Universe is adiabatic and reversible so that a polytropic relation holds,
$p=\tilde p \rho ^\gamma$ \cite{octo}, where $\gamma=C_p/C_V$.
It is well known that currently the perturbations of the density, $\delta\rho$, barely propagate in the Universe \cite{sound}. It then follows that the speed of sound
$c_s^2 = \left(\frac{\partial p}{\partial \rho}\right)$
can be considered as vanishing. On the other hand, from the above thermodynamic assumptions we obtain for the
Universe  $c_s= \sqrt{\gamma (C_p-C_V)T}$ so that for a vanishing sound speed
the allowed solutions are formally $C_p-C_V=0$ or $\gamma=0$. The first solution implies a pressureless universe whereas the
second solution leads to a heat capacity $C_p=0$, equivalent to $h=const$, and a pressure $p=-\rho C_V T$ which is negative for positive values
of the heat capacity $C_V$.

We conclude that the important result of assuming an ideal-gas-like universe with zero speed of sound
is that the pressure in that universe can be zero or negative\footnote{Similar results were obtained by analyzing
the ideal gas in the context of  geometrothermodynamics \cite{vqs10}}.
Moreover, let us notice that this is a consequence of the fact
that each particle of a given fluid
undergoes an early isentropic process, when the wave amplitude is infinitesimal. This condition follows
from the fact
that, in general, the entropy is proportional to the square of the
velocity and temperature gradients. 
We will now consider a vanishing speed of sound in the context of a FRW cosmology. Without putting any
further information into the Einstein equations, we expect as a result
a cosmological model in which the total pressure is negative and constant.
Considering for the sake of simplicity the redshift $z$ as the ''time" variable defined by $dz/dt=-(1+z)H$, where $t$ is the
time coordinate, the conservation law for a generic fluid with EoS
$p=w(z)\rho$, is
\be\label{cons}
\frac{d\rho}{dz}-\frac{3(1+w)}{1+z}\rho=0\,.
\ee
Assuming $c_s^2=\partial p/\partial\rho=0$, we get
\be
\frac{d w}{dz}+\frac{3w(1+w)}{1+z}=0\,,
\ee
whose solutions are
\begin{equation}\label{bz}
w=0 \quad {\rm and} \quad
w=-\frac{1}{1-\xi(1+z)^3}\,.
\end{equation}
Notice that as a consequence of the thermodynamic hypothesis presented above, we obtain in a straightforward manner a dust-like term and an additional term with a time--dependent barotropic factor. This is a result of our model which is otherwise usually
postulated arbitrarily in cosmology \cite{duat}.

Introducing  the above solutions for $w$ into Eq. ($\ref{cons}$), we obtain the most general density for \emph{one fluid} satisfying
the condition $c_s=0$
\be
\rho(z)= (\rho_m+\tilde\rho \xi)(1+z)^3 - \tilde \rho\,.
\label{rho}
\ee
Notice that the term $\rho_m(1+z)^3$ corresponds to the solution $w=0$. This general solution involves three constant
parameters, namely, $\xi$, $\rho_m$ and $\tilde\rho$.

Introducing $\rho(z)$ into the first Friedmann equation $H^2= (8\pi G/3)\rho(z)$, we obtain the generic normalized Hubble rate, $E\equiv\frac{H}{H_0}$,
\begin{equation}\label{Hz}
E=\sqrt{\left(\Omega_X+1\right)(1+z)^3-\Omega_X}\,,
\end{equation}
where we adopt the convention $\Omega_X\equiv\frac{\tilde{\rho}}{\rho_c}$, with the critical density defined as $\rho_c\equiv\frac{3H_{0}^{2}}{8\pi G}$. In Eq. ($\ref{Hz}$), we used the condition $E(z=0)=1$, which gives
\begin{equation}\label{fr}
\xi=\frac{1+\Omega_X-\Omega_m}{\Omega_X}\,,
\end{equation}
where $\Omega_m\equiv\frac{\rho_m}{\rho_c}$. Therefore, to determine the Hubble rate $E(z)$ we need the constant $\Omega_X$,
whereas to determine the barotropic factor $w(z)$ in Eq. (\ref{bz})  it is necessary to know the two constants
$\Omega_m$ and $\Omega_X$.

Using the relationship  as $\Omega_X=\frac{\Omega_m-1}{1-\xi}$ which follows from Eq. ($\ref{fr}$),
it is possible to rewrite  Eq. ({\ref{Hz}}) as
\begin{equation}\label{Hz2}
E=\sqrt{\tilde\Omega_m(1+z)^3+\tilde\Omega_\Lambda}\,,
\end{equation}
where  $\tilde\Omega_m\equiv\frac{\xi-\Omega_m}{\xi-1}$ and the cosmological constant term reads $\tilde\Omega_\Lambda\equiv1-\tilde\Omega_m$.

The $\Lambda$CDM model is contained in Eq. (\ref{rho}) in the limiting case $\xi=0$ with $\tilde \rho=\rho_m -1$ so that
$\rho_m$ turns out to represent the sum of the
baryonic and the cold dark matter densities.
It follows that the model presented here is
a generalization of  $\Lambda$CDM and it is the result of the physical assumption that the Universe is made of only one matter fluid in which the the speed of sound is required to vanish.
Hence, the crucial difference lies in the fact that in the present model no cosmological constant is postulated \emph{a priori}.

In addition, difficulties related to  the well-known problems of coincidence
and of fine tuning \cite{copeland} are solved in the context of the present model by
the presence of a variable barotropic factor $w(z)$ and by the fact that no cosmological
constant is assumed to be related to the vacuum energy.

For instance, if we consider $\Omega_m=0.274$ and $w=-0.980$ \cite{kow}, we get from Eqs. ($\ref{bz}$) and ($\ref{fr}$) $\xi\approx-0.02$ and $\Omega_X\approx-0.712$, which represents a value of $\Omega_X$, compatible with an expanding Universe.

Moreover, it is possible to infer the limits of the evolution of $w(z)$ as
\begin{equation}\label{w0in}
\left\{
  \begin{array}{lll}
    w_0=\frac{1}{\xi-1}, & \hbox{$for\quad z=0$\,;} \\
\\
    w_\infty=0, & \hbox{$for\quad z\rightarrow\infty$\,.}
  \end{array}
\right.
\end{equation}
Eqs. ($\ref{w0in}$) show that at redshift $z=0$ $w_0$ is a constant which predicts $\xi\leq0$ for $w\geq-1$.
At higher redshift, the usual dust--like component dominates, because $w_\infty=0$.

If the Universe accelerates, the so-called acceleration parameter, defined as
\begin{equation}\label{qz}
q=-1-\frac{\dot H}{H^2}\,,
\end{equation}
must be negative.
From Eq. ($\ref{qz}$), for our model we obtain
\begin{equation}\label{jd}
q=-1+\frac{3\left(1+\Omega_X\right)(1+z)^3}{2+2\left(1+\Omega_X\right)z\Big[3+z(3+z)\Big]}\,,
\end{equation}
so that at $z=0$ it reduces to
\begin{equation}\label{q0}
q_0=\frac{1}{2}\left(1+3\Omega_X\right)\,.
\end{equation}
For the particular value $\Omega_X=-0.712$, we get $q_0\approx-0.57$ which is in agreement with observations \cite{visser}.

At the moment in which the acceleration starts ($q=0$), the correspondent redshift reads
\begin{equation}\label{red}
z_{acc}=-1+\frac{\Big[2(-\Omega_X-2\Omega_X^2-\Omega_X^3)\Big]^{\frac{1}{3}}}{1+\Omega_X}\,,
\end{equation}
so that for $\Omega_X=-0.712$ we have $z_{acc}\approx0.7$.

\begin{figure}[h]
  \includegraphics[scale=0.8]{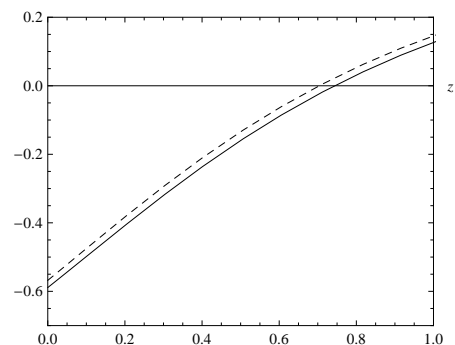}\\
  \caption{In this graphic is plotted $q(z)$ for our model (dashed line) and $\Lambda$CDM (black line). The indicative values are $\Omega_m=0.274$, $\Omega_X=-0.712$.}\label{uno}
\end{figure}

Using Eqs. ($\ref{Hz}$) and ($\ref{Hz2}$), it is possible to perform an experimental procedure to constrain the values of the
constants $\Omega_m$, $\Omega_X$ and $\xi$. In
particular, we employ the three most common fitting procedures:
Supernovae Ia (SNeIa), Baryonic Acoustic Oscillation (BAO) and Cosmic Microwave Background (CMB).
We will use of the most recent updated
Union 2 compilation \cite{kow}, which alleviates the problem of systematics.

Thus, associating to each Supernova modulus $\mu$ the
corresponding $1\sigma$ error, denoted by $\sigma_{\mu}$, we define the distance modulus $\mu = 25 + 5 \log_{10} \frac{d_L}{Mpc}$,
where $d_L(z)$ is the luminosity distance
\begin{equation}
d_L(z) =  (1+z) \int_0^z \frac{dz'}{H(z')}\,,
\end{equation}
and we minimize the chi square, defined as follows
\begin{equation}
\chi^{2}_{SN} =
\sum_{i}\frac{(\mu_{i}^{\mathrm{theor}}-\mu_{i}^{\mathrm{obs}})^{2}}
{\sigma_{i}^{2}}\,.
\end{equation}

The second test that we perform is related to the observations of
large scale galaxy clusterings, which provide the signatures of the
BAO \cite{Percival:2009xn}. We use the measurement of
the peak of luminous red galaxies observed in Sloan Digital Sky
Survey (SDSS), denoted by $A$
\begin{equation}
A=\sqrt{\Omega_m}  \Big[\frac{H_0}{H(z_{BAO})}\Big]^{\frac{1}{3}}
\left[ \frac{1}{z_{BAO}}\int_0^{z_{BAO}}
\frac{H_0}{H(z)}dz\right]^{\frac{2}{3}}\,,
\end{equation}
with $z_{BAO}=0.35$. In addition, the observed $A$ is estimated to
be
\begin{equation}\label{ax}
A_{obs} = 0.469 \left(\frac{0.95}{0.98}\right)^{-0.35}\,,
\end{equation}
with an error $\sigma_A = 0.017$. In the case of the BAO measurement
we minimize the chi square
\begin{equation}
\chi^{2}_{BAO}=\left(\frac{A-A_{obs}}{\sigma_A}\right)^2\,.
\end{equation}

Finally, for the CMB test we define the so-called CMB
shift parameter
\begin{equation}
R=\sqrt{\Omega_m} \int_0^{z_{CMB}} \frac{H_0}{H(z)} dz\,,
\end{equation}
with $z_{CMB}=1091.36$  \cite{Wang:2007mza}. It gives a
complementary bound to the SNeIa data and BAO because the SNeIa redshift is $z<2$, $z_{BAO}=0.35$, while here $z\sim 1100$.

We minimize the chi square
\begin{equation}
\chi^{2}_{CMB}=\left(\frac{R-R_{obs}}{\sigma_R}\right)^2\,.
\end{equation}

It is important to note that BAO and CMB do not depend on the values of $H_0$.
We summarize the results of this numerical procedure  in Tab. I

\begin{table}[h]
\begin{center}\label{I}
\begin{tabular}{|c|c|c|c|}
\hline
$\Omega_{m}(SN)$  &  $\Omega_{m}(BAO)$ & $\Omega_{m}(CMB)$ \\
\hline
$0.275\pm0.016$ & $0.264\pm0.016$ & $0.280\pm0.019$ \\
\hline
$\xi(SN)$  &  $\xi(BAO)$ & $\xi(CMB)$ \\
\hline
$0.036\pm0.009$ & $-0.050\pm0.011$ & $-0.040\pm0.012$\\
\hline
$\Omega_X(SN)$  &  $\Omega_X(BAO)$ & $\Omega_X(CMB)$ \\
\hline
$-0.765\pm0.058$ & $-0.700\pm0.061$ & $-0.729\pm0.050$\\
\hline
\end{tabular}
\caption{Summary of the numerical results for our model; the constants $\Omega_m$ and $\xi$ have been fitted by using Eq. ($\ref{Hz2}$) and the parameter $\Omega_X$ by using Eq. ($\ref{Hz}$). The chi square's are
$\chi^{2}_{SNeIa}=1.010$, $\chi^{2}_{BAO}=0.989$,
$\chi^{2}_{CMB}=1.000$, and $\chi^{2}_{SNeIa}=1.030$, $\chi^{2}_{BAO}=1.010$,
$\chi^{2}_{CMB}=0.997$, respectively for the test with Eq. (\ref{Hz}) and Eq. ($\ref{Hz2}$); the mean values for the three parameters are $\Omega_{m\,(mean)}=0.273$,
$\Omega_{X\,(mean)}=-0.731$ and $\xi_{mean}=-0.042$. Note that the value of $H_0$ for the SNeIa tests is $H_0=72\pm5\, Km\,s^{-1}\,Mpc^{-1}$.}
\end{center}
\end{table}

The values of $\Omega_m$, $\Omega_X$ and $\xi$ of Tab. I are in agreement with the theoretical results showed previously.

During the last decades, different parametrizations of $w(z)$ were proposed \cite{copeland}. For instance,  $w=w_1+w_2z$, $w=w_\alpha+w_\beta\log(1+z)$ or $w=w_0+w_a(1-a)$. In particular, the third case was introduced by Chevallier,
Polarski and Linder and it is referred to as the CPL
parametrization \cite{CPL}. The CPL parametrization has the
advantages that at low and very high redshift it reduces to constant
values, respectively $w(z\rightarrow0)=w_0$, and
$w(z\rightarrow\infty)=w_0+w_a$. However, all the  parametrizations suggested so far for
$w(z)$ are either  {\it ad hoc} proposals or the result of phenomenological assumptions only. Our model predicts
a theoretical barotropic factor $w(z)$, and it is also able to reproduce previous results (see, for instance, \cite{sound}).

Moreover, the barotropic factor is also connected to an interesting quantity, that one might consider as a natural measure of time variation,
namely, $\frac{dw}{d \ln(1 + z)}\Big|_{z=1}$. In models involving a scalar field $\varphi$ with potential $V(\varphi)$,
this quantity is related to the slow-roll potential
$\propto\frac{V^{'}}{V}$, in the region $z = 1$, where the scalar
field is most likely to be evolving as the epoch of matter
domination changes over to dark energy\footnote{It is a common opinion to refer to the missing ingredient, driving the acceleration, as dark energy.}
domination. For the CPL parametrization $\frac{dw}{d \ln(1 + z)}\Big|_{z=1}=\frac{w_a}{2}$, while for the present model it is
\begin{eqnarray}\label{ns}
\frac{dw}{d \ln(1 + z)}\Big|_{z=1} =-24\Omega_X\frac{1+\Omega_X-\Omega_m}{\left(8+7\Omega_X-8\Omega_m\right)^2}\,.
\end{eqnarray}
By comparing our result with CPL we have two solutions. One of these solutions is physically compatible with the accelerating scenario; in fact,
by considering $\Omega_{m (mean)}$ and the indicative value $w_a=0.58$ \cite{CPL}, we find $\Omega_X\approx-0.716$, in agreement with the observational results.

We present below the graphics of the evolution of $w(z)$, $q(z)$ and the expansion history for $a(t)$,
given by \cite{CPL,expansion}
\begin{equation}\label{expan}
H_0t(a)=\int_{a}^{1}\frac{da^{'}}{a^{'}\,E\left(a^{'}\right)}\,.
\end{equation}

\begin{figure}[h]
  \includegraphics[scale=0.8]{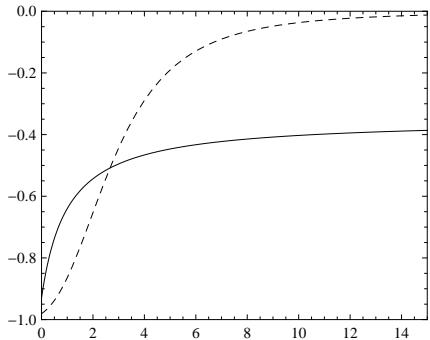}\\
  \caption{In this graphic is plotted $w(z)$ (Y axis) for our model (dashed line) and CPL (black line). The indicative values are $\Omega_m=0.274$, $\Omega_X=-0.712$, $w_0=-0.93$ and $w_a=0.58$.}\label{due}
\end{figure}

\begin{figure}[h]
  \includegraphics[scale=0.8]{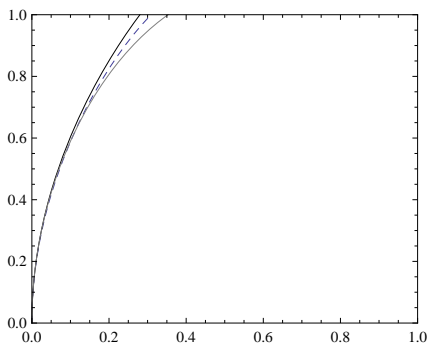}\\
  \caption{In this graphic is plotted the expansion history of $a(t)$ (X axis) versus $H_0t$ (Y axis) for our model (dashed line), $\Lambda$CDM (black line) and CPL (grey line). The indicative values are $\Omega_m=0.274$, $\Omega_X=-0.712$, $w_0=-0.93$ and $w_a=0.58$.}\label{tre}
\end{figure}

The presence of baryonic and dark matter is generally intertwined
with the addition of an exotic fluid which drives the acceleration.
Unfortunately, all the attempts to describe this unexpected acceleration, suffer from various shortcomings. Moreover, the $\Lambda$CDM remains the favorite fitting model to describe the Universe dynamics, by including in Einstein equations a second fluid characterized
by the cosmological constant. We showed that it is possible to discard the existence of a second fluid and, instead, to
use ordinary thermodynamics. Assuming a vanishing speed of sound in order to guarantee that small matter perturbations do not propagate, we obtain
a theoretical parametrization
for $w$, which generalizes the $\Lambda$CDM model, reducing to it in a special case.
In addition, the model presented here  solves the coincidence and fine tuning problems in a straightforward manner.
Using the cosmological tests of SNeIa, BAO and CMB, it was shown that our model is able to reproduce the observable Universe, and is in
agreement with the theoretical limits.

\section*{Acknowledgements}

One of the authors (O.L.) is grateful to B. Luongo and G. Capasso. This work was supported in part by DGAPA-UNAM, grant No. IN106110.


\begin{thebibliography}{99}

\bibitem{SNeIa}
Riess, A. G.,  et al., AJ, {\bf 116}, 1009, (1998); Perlmutter, S., et al., ApJ,
{\bf 517}, 565, (1999).

\bibitem{copeland}
Copeland, J. E., Sami, M., Tsujikawa, S., Int. J. Mod. Phys. D,
{\bf 15}, 1753-1936, (2006).

\bibitem{au}
Padmanabhan, T., Phys. Rept., {\bf 380}, 235, (2003); Tsujikawa, S., ArXiv: 1004.1493, (2010).

\bibitem{Lambda}
Sahni, V., Starobinski, A., Int. J. Mod. Phys. D, {\bf 9}, 373, (2000); Tegmark, M., et al., Phys. Rev. D, {\bf 69}, 103501, (2003).

\bibitem{sound}
Kunz, M., Phys. Rev. D, {\bf 80}, 123001, (2009).

\bibitem{other}
Weinberg, S., Rev. Mod. Phys., {\bf 61}, 1, (1989).

\bibitem{kqs}
Krasinski, A., Quevedo, H.,  and Sussman, R., J. Math. Phys., {\bf 38}, 2602, (1997).

\bibitem{octo}
Kundu, P. K., Cohen, I. M., Fluid Mechanics, Elsevier Acad. Press, San Diego, USA, (2004).

\bibitem{vqs10}
Vazquez, A., Quevedo, H., Sanchez, A., J. Geom. Phys., {\bf 60}, 1942, (2010).

\bibitem{duat}
Hu, W., Eisenstein, D. J., Phys. Rev. D, {\bf 59}, 083509, (1999); Wassermann, I., Phys. Rev. D, {\bf 66}, 123511, (2002); Rubano, C., Scudellaro, P., Gen. Rel. Grav., {\bf 34}, 1931, (2002).

\bibitem{kow}
Komatsu, E., et al., Astrophys. J. Suppl., {\bf 192}, 18, (2011).

\bibitem{visser}
Visser, M., $\&$ Catto$\ddot{e}$n, C., Class. Quant. Grav., {\bf 24}, 5985, (2007); Cattoen, C., Visser, M., Phys. Rev. D, {\bf 78}, 063501, (2008).

\bibitem{Percival:2009xn}
Percival, W. J., et al., Mon. Not. Roy. Astron. Soc., {\bf 401}, 2148,
(2010).

\bibitem{Wang:2007mza}
Eisenstein, D. J., et al, Astrophys. J., {\bf 633}, 560, (2005).

\bibitem{CPL}
Chevallier, M., Polarski, D., Int. J. Mod. Phys. D., {\bf 10}, 213, (2001);
Linder, E., Phys. Rev. Lett., {\bf 90}, 091301, (2003).

\bibitem{expansion}
Weinberg, S., Cosmology, Oxford Univ. Press, New York, USA, (2008).

\end{thebibliography}
\end{document}